\documentclass[aps,preprint]{revtex4}
\usepackage{amsmath,amssymb,amsfonts,graphicx,hyperref,amsthm}

\begin{document}

\title{Casimir Effect in the Kerr Spacetime with Quintessence}

\author{V. B. Bezerra}
%\email{valdir@fisica.ufc.br}
\address{Departamento de F\'isica, Universidade Federal da Para\'iba, Caixa Postal 5008, Jo\~ao Pessoa-PB, 58051-970, Brazil.}

\author{M. S. Cunha }
%\email{marcony.cunha@uece.br}
\address{Grupo de F\'isica Te\'orica (GFT), Centro de Ci\^encias e Tecnologia, Universidade Estadual do Cear\'a, CEP 60740-000, Fortaleza, Cear\'a, Brazil.}

\author{L. F. F. Freitas}
%\email{felipefreitas@fisica.ufc.br}
\address{Departamento de F\'sica, Universidade Federal do Cear\'a, Caixa Postal 6030, CEP 60470-455, Fortaleza, Cear\'a, Brazil.}

\author{C. R. Muniz}
\email{celio.muniz@uece.br, corresponding author.}\address{Grupo de F\'isica Te\'orica (GFT), Universidade Estadual do Cear\'a, Faculdade de Educa\c c\~ao, Ci\^encias e Letras de Iguatu, CEP 63500-000, Iguatu, Cear\'a, Brazil.}

\author{M. O. Tahim}
%\email{makarius.tahim@uece.br}
\address{Grupo de F\'isica Te\'orica (GFT), Universidade Estadual do Cear\'a, Faculdade de Educa\c c\~ao, Ci\^encias e Letras do Sertão Central, CEP 63900-000, Quixad\'a, Cear\'a, Brazil.}

\begin{abstract}

We calculate the Casimir energy of a massless scalar field in a cavity formed by nearby parallel plates orbiting a rotating spherical body surrounded by quintessence, investigating the influence of the gravitational field on that energy, at zero temperature. This influence includes the effects due to the spacetime dragging caused by the source rotation as well as those ones due to the quintessence. We show that the energy depends on all the involved parameters, as source mass, angular momentum and quintessence state parameter, for any radial coordinate and polar angle. We show that at the north pole the Casimir energy is not influenced by the quintessential matter. At the equatorial plane, when the quintessence is canceled, the result obtained in the literature is recovered. Finally, constraints in the quintessence parameters are obtained from the uncertainty in the current measurements of Casimir effect.\\

\vspace{0.75cm}
\noindent{Key words: Casimir effect, Kerr spacetime, quintessence.}
\end{abstract}

\maketitle

\section{Introduction}

The actual theoretical and experimental researches concerning the structure and evolution of the Universe involve knowing what is its basic stuff. Besides the barionic and dark matters, recent investigations point to a kind of cosmic energy responsible for the accelerated expansion of the Universe observed since 1998, whose origin and nature is still unknown and which was measured recently with reasonable accuracy \cite{PlanckSat}. Moreover, such an exotic energy, also called dark energy, should be the principal constituent of the Universe in order to account for its observed spatial flatness. There are models as the $\Lambda$CDM based on a cosmological constant, where dark energy has a constant density throughout the Universe, as well as there are those ones which consider a spatially and temporally variable density of that energy, named quintessence \cite{Caldwell1,Doran} (and references therein).

Quintessence behaves like the inflaton field studied in the inflationary cosmology, which is often modeled as a scalar field slowly rolling down a potential barrier. The approaches which make use of such a field yield results completely different from a description via cosmological constant, since a dynamical field presents wavelength fluctuations which are unrevealed in the cosmic microwave background (CMB) through its temperature variations \cite{Weinberg}, which were precisely measured by the Planck satellite \cite{PlanckSat}, as well as in the large scale mass distribution, which would involve influences on clustering dark matter at remote ages \cite{Caldwell2}. Besides this, quintessence models based on supergravity or string theories accounts to the transition of the Universe for its current accelerated phase \cite{Gardner}. On the other hand, at astrophysical scales, we can presume the existence of quintessential matter concentrated around of stars and compact objects, which could cause remarkable effects as the gravitational shift of the light coming from distant stars \cite{Liu}. These effects can be more appropriately studied in the context of the general relativity by means of some its exact solutions, as for example, the ones corresponding to black holes surrounded by quintessence \cite{Kiselev,Ghosh,Xu}.

A way of locally evaluating the presence of this stuff in our spatial environment is by means of the Casimir effect, as it already was made in order to study the cosmological constant \cite{Leandros}. The Casimir effect \cite{Casimir} is a physical phenomenon which was originally predicted as the attractive force arising between two parallel, uncharged metallic plates in perfect vacuum. As there is no force between uncharged plates in classical electrodynamics, it is possible to conclude that this effect is of quantum origin and results from the modifications of the zero-point oscillations of the electromagnetic field by the material boundaries. This phenomenon appears for other quantum fields and for material boundaries made of various materials, kept at temperatures different from zero \cite{Trunov,Milton,BORDAG,Banishev}.

The Casimir effect also arises in empty spacetimes with nontrivial topology. In this case, there are no material boundaries. These are substituted by some identification conditions which are imposed on the quantum fields. Thus, the Casimir effect gives rise to a nonzero stress tensor of a quantum field in a vacuum state which depends on the geometrical and topological features of the manifold. As an example we can mention the static Einstein universe with the topology $S^{3} \times R$, where the Casimir energy density and pressure for a massles scalar field were found some time ago \cite{larry,herondy1,herondy2}. Following these line of research, the Casimir effect was investigated in a wide class of spacetimes with nontrivial topology and with several kind of metrics, including alternative theories of gravity \cite{Elizalde1,Elizalde2,Elizalde3,Galina,Sorge1,Sorge2,Zhang,Celio1,Celio2,Celio3}.

In this paper, we calculate the Casimir energy of a massless scalar field in a cavity formed by nearby parallel plates orbiting a rotating spherical body surrounded by quintessence, at zero temperature, investigating the effects generated by both the gravitational field and the dragging of spacetime around the body, as well as by the quintessence on the regularized vacuum energy of the scalar field. The study follows that one made in \cite{Sorge2}, which analyzed the gravitational and rotational effects due to Kerr spacetime on that energy. The organization of the paper is as follows: Section II is devoted to present briefly the spacetime which will be considered. Section III details the solution of the Klein-Gordon equation. In Section IV, the calculation of the Casimir energy at zero temperature is presented. Finally, Section V exposes the conclusions and remarks.

\section{Kerr black hole surrounded by Quintessence}

The spacetime generated by an axially symmetric gravitational source of mass $M$ and angular momentum $J$, surrounded by quintessence characterized by a state parameter $\omega_o$, which is the ratio between its mass density and pressure, $\rho_q/p_q=\omega_o$, is described by the metric recently found by S. G. Ghosh \cite{Ghosh}, which was obtained from Newman-Janis complex transformations on the static counterpart of a black hole with quintessence \cite{Kiselev}. Ghosh's metric is given, in Boyer-Linquidist coordinates, by the expression
\begin{equation}
ds^2=\left(\frac{\widetilde{\Delta}-a^2\sin^2{\theta}}{\Sigma}\right)\!dt^2+2a\left(1-\frac{\widetilde{\Delta}-a^2\sin^2{\theta}}{\Sigma}\right)\sin^2{\theta} dtd\phi-\frac{\Sigma}{\widetilde{\Delta}} dr^2-\Sigma d\theta^2-\frac{\widetilde{A}}{\Sigma}\sin^2{\theta} d\phi^2, \label{KerrQuintMetr}
\end{equation}
where
\begin{eqnarray}
&\Sigma=r^2+a^2\cos^2\!\theta, & \nonumber \\
&\widetilde{\Delta}=r^2+a^2-2Mr-\overline{\alpha}\Sigma^{(1-3\omega_{o})/2}, & \\  \label{KerrParam}
&\widetilde{A}=\Sigma^2+a^2\sin^2{\theta}\left(2\Sigma-\widetilde{\Delta}+a^2\sin^2{\theta}\right), & \nonumber
\end{eqnarray}
In this metric, $a=J/M$ is the angular momentum per mass unit and $\omega_o$ is the quintessential state parameter which runs in the interval $[-1,-1/3]$ when one considers the actual accelerated expansion of the Universe pushed by dark energy. The special values $\omega_o=1/3$, $\omega_o=0$ and $\omega_o=-1$ correspond to universes with dominance of radiation, matter, and cosmological constant, respectively. The values $\omega_o<-1$ lead to the so-called phantom energy \cite{Caldwell}. The parameter $\overline{\alpha}$ is such that the state equation for the quintessential mass density is given by
\begin{equation}\label{quintedensity}
\rho_q=-\frac{3\overline{\alpha}\omega_o}{2r^{3(\omega_o+1)}},
\end{equation}
and the pressure is
\begin{equation}\label{quintpress}
p_q=-\frac{3\overline{\alpha}}{2r^{3(\omega_o+1)}}.
\end{equation}
For $\overline{\alpha}=0$ Eq. (\ref{KerrQuintMetr}) reduces to the standard Kerr black hole (with $\widetilde{\Delta}=\Delta$ and $\widetilde{A}=A$). In the particular case $a=0$, the same metric yields the Schwarzschild solution surrounded by the quintessence, according to the Kiselev description \cite{Kiselev}. The metric (\ref{KerrQuintMetr}) is then a rotating quintessence black hole which is stationary and axisymmetric with Killing vectors.

Here, we will extend the work of Sorge \cite{Sorge2} which particularized the problem to the Kerr metric in the equatorial plane. For convenience, and without loss of generality, we consider the metric approximately constant between the plates, since the separation distance between them, $L$, is much smaller than the radius of the orbit, ($L \ll r$). The cavity is in a locally co-moving referential system which has Cartesian coordinates $(x, y, z)$ defined on one of the plates, so that the $z$ axis is tangential to the path of the circular orbit (see Fig.1 below).
\begin{figure}[h!]
\centering
\includegraphics[scale=0.5]{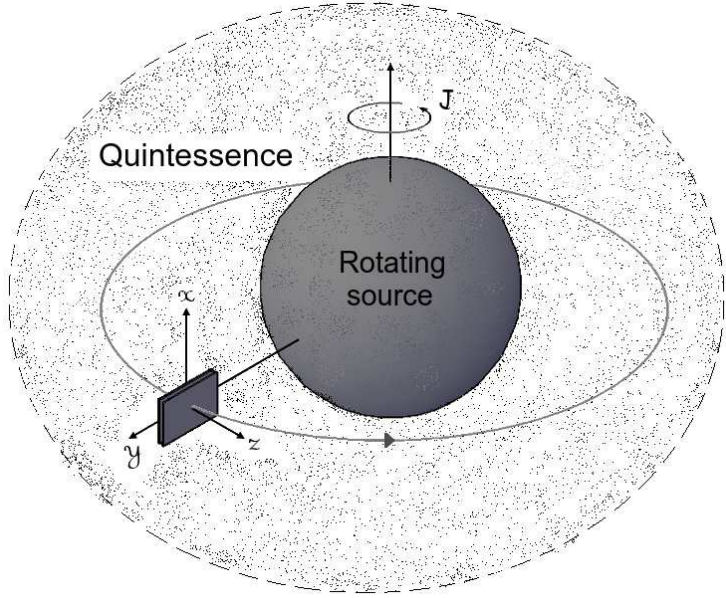}
\caption{Representation of a Casimir cavity with a Cartesian reference frame orbiting around a rotating gravitational source surrounded by quintessence.}
\end{figure}

Thus, the relationship between the spherical coordinates of the system orbiting the source with angular velocity $\Omega$ and the Cartesian one in the plates is $ dy = dr$, $ dz = rd\phi' $ and $ dx = -rd\theta $, where $\phi'=\phi-\Omega t$. Therefore, in this Cartesian system the metric becomes
\begin{equation}
ds^2 =\widetilde{C}^{-2}\left(\Omega\right)\!dt^2+2\frac{\widetilde{A}}{r \Sigma}\sin^2\!\theta\left(\omega_{d}-\Omega\right) dtdz-\frac{\Sigma}{\widetilde{\Delta}} dy^2-\frac{\Sigma}{r^{2}} dx^2-\frac{\widetilde{A}}{r^{2}\Sigma}\sin^2\!\theta dz^2, \label{01}
\end{equation}
and $\widetilde{\omega}_{d}$ is the angular velocity of the dragging of the spacetime around the gravitational source, namely
\begin{equation}
\widetilde{\omega}_{d}=-\frac{g_{t\phi}}{g_{\phi\phi}}=\frac{2Mar}{\widetilde{A}}.\label{Eq.12}
\end{equation}
Besides this, we get
\begin{equation}
\widetilde{C}^{-2}\left(\Omega\right)=\frac{\Sigma \widetilde{\Delta}}{\widetilde{A}}\left[1-\frac{\widetilde{A}^{2}}{\widetilde{\Delta} \Sigma ^2}\sin^2{\theta}\left(\Omega-\widetilde{\omega}_d\right)^2\right]. \label{03}
\end{equation}
With these equations, we can calculate the normal modes and the Casimir energy. We call attention again to the fact that by considering $\overline{\alpha}=0$ (without quintessence, which removes the tilde signal of the above expressions) and making $\theta=\pi/2$ we back to the original problem solved in \cite{Sorge2}.

\section{Normal modes for a massless scalar field}

The massless scalar field is confined in a cavity which orbits the black hole as showed in Fig.(01). This filed obeys the covariant Klein-Gordon equation, given by
\begin{equation}
\left[\frac{1}{\sqrt{-\hat{g}}}\partial_{\mu}\left(\sqrt{-\hat{g}}\hat{g}^{\mu \nu}\partial_{\nu}\right)+\xi R\right]\phi\left(t, \textbf{x} \right)=0, \label{04}
\end{equation}
where we will consider $\xi = 0$ (minimum coupling). As said before, we also consider the metric (\ref {01}) approximately constant between the plates, since the separation distance between them is much smaller than the radius of the orbit ($L\ll r$). Thus, the equation (\ref {04}) takes the form,
\begin{equation}
\hat{g}^{\mu\nu}\partial_{\mu}\partial_{\nu}\!\phi\left(t,\mathbf{r}\right)=0, \label{05}
\end{equation}
where $\hat{g}^{\mu\nu}$ is the inverse metric of Eq. (\ref{01})
\begin{equation}
\left(\hat{g}^{\mu\nu}\right)=\left[\begin{array}{cccc}
\frac{\widetilde{A}}{\Sigma\widetilde{\Delta}}& 0 & 0 & -\frac{r \widetilde{A}\left(\Omega-\widetilde{\omega}_{d}\right)}{\widetilde{\Delta} \Sigma} \\
0 & -\frac{r^{2}}{\Sigma} & 0 & 0  \\
0 & 0 & -\frac{\widetilde{\Delta}}{\Sigma} & 0 \\
-\frac{r \widetilde{A}\left(\Omega-\widetilde{\omega}_{d}\right)}{\widetilde{\Delta} \Sigma} & 0 & 0 & - \frac{r^{2}}{\sin^2\!\theta\widetilde{\Delta}}\widetilde{C}^{-2}_{\Omega}
\end{array}\right]. \label{06.1}
\end{equation}
Proposing a solution to Eq. (\ref{05}) in the form $\phi\left(t,\mathbf{r}\right)\propto e^{i\left(k_{x}x+k_{y}y-\omega t\right)}Z\left(z\right)$ we get
\begin{equation}
\hat{g}^{zz}\frac{d^{2}}{dz^{2}}Z\!\left(z\right)-2i\omega\hat{g}^{tz}\frac{d}{dz}Z\!\left(z\right)-\left(k_{x}^{2}\hat{g}^{xx}+k_{y}^{2}\hat{g}^{yy}+\omega^{2}\hat{g}^{tt}\right)\!\!Z\!\left(z\right)=0, \label{07}
\end{equation}
where it is supposed a solution of the type $Z\!\left(z\right)=e^{i \alpha z}$, and then
\begin{equation}
\hat{g}^{zz}\alpha^{2}-2\omega\hat{g}^{tz}\alpha+\left(k_{x}^{2}\hat{g}^{xx}+k_{y}^{2}\hat{g}^{yy}+\omega^{2}\hat{g}^{tt}\right)=0. \label{08}
\end{equation}
Thus, the solutions (\ref{08}) for $\alpha$ are given by
\begin{equation}
\alpha=\frac{\omega\hat{g}^{tz}}{\hat{g}^{zz}}\pm\sqrt{\left(\frac{\omega\hat{g}^{tz}}{\hat{g}^{zz}}\right)^{2}-\left(\frac{k_{x}^{2}\hat{g}^{xx}+k_{y}^{2}\hat{g}^{yy}+\omega^{2}\hat{g}^{tt}}{\hat{g}^{zz}}\right)}. \label{09}
\end{equation}
Substituting the values of $\hat{g}^{\mu\nu}$, Eq. (\ref{09}) results in
\begin{equation}
\alpha=\frac{\omega \widetilde{A} \left(\Omega-\widetilde{\omega}_{d}\right)\sin^{2}{\theta} \widetilde{C}^{2}_{\Omega}}{r\Sigma}\pm\frac{\sqrt{\widetilde{\Delta}}\sin{\theta} \widetilde{C}^{2}_{\Omega}}{r}\sqrt{\omega^{2}-\widetilde{C}^{-2}_{\Omega}\frac{r^2}{\Sigma}\left(k_{x}^{2}-\frac{\widetilde{\Delta}}{r^{2}}k_{y}^{2}\right)}, \label{10}
\end{equation}
and the solution for $Z\!\left(z\right)$ can be written therefore as
\begin{equation}
Z\!\left(z\right)=\exp{\left[i\frac{\omega \widetilde{A} \left(\Omega-\widetilde{\omega}_{d}\right)\sin^{2}{\theta} \widetilde{C}^{2}_{\Omega}}{r\Sigma}z\right]}\sin{\left[z\frac{\sqrt{\widetilde{\Delta}}\sin{\theta} \widetilde{C}^{2}_{\Omega}}{r}\sqrt{\omega^{2}-\widetilde{C}^{-2}_{\Omega}\frac{r^2}{\Sigma}\left(k_{x}^{2}-\frac{\widetilde{\Delta}}{r^{2}}k_{y}^{2}\right)}\right]}. \label{11}
\end{equation}

The solutions of Eq. (\ref{05}) must satisfy the Dirichlet boundary conditions on the plates. Considering one of them at $z=0$  and the other at $z=L$, we get $\phi\left(t,x,y,0\right)=\phi\left(t,x,y,L\right)=0$. From these conditions we conclude that
\begin{equation}
\sin{\left[z\frac{\sqrt{\widetilde{\Delta}}\sin{\theta} \widetilde{C}^{2}_{\Omega}}{r}\sqrt{\omega^{2}-\widetilde{C}^{-2}_{\Omega}\frac{r^2}{\Sigma}\left(k_{x}^{2}-\frac{\widetilde{\Delta}}{r^{2}}k_{y}^{2}\right)}\right]}=0, \label{12}
\end{equation}
obtaining thus the eigenfrequencies
\begin{equation}
\omega_{n}=\frac{r}{\sqrt{\widetilde{\Delta}}\sin\!\left(\theta\right)  \widetilde{C}^{2}_{\Omega}}\left[\left(\frac{n\pi}{L}\right)^{2}+\frac{\widetilde{\Delta}\sin^2{\theta}  \widetilde{C}^{2}_{\Omega}}{\Sigma}\left(k_{x}^{2}+\frac{\widetilde{\Delta}}{r^{2}}k_{y}^{2}\right)\right]^{\frac{1}{2}}. \label{14}
\end{equation}
Writing the eigenfunctions for the scalar field, $\phi_n$, we have
\begin{equation}
\phi_{n}\left(t,x,y,z\right)=N_{n}e^{i\left(k_{x}x+k_{y}y-\omega_n t\right)}e^{i\beta_{n}z}\sin{\!\left(\frac{n \pi}{L}z\right)}, \label{15}
\end{equation}
where $\beta_{n}=\frac{\widetilde{A}\left(\Omega-\widetilde{\omega}_{d}\right)\sin^{2}{\theta} \widetilde{C}^{2}_{\Omega}}{r\Sigma}\omega_n$ and $N_{n}$ is the normalization constant, which can be obtained from the internal product $\langle\phi_{n},\phi_{n}\rangle=||\phi_{n}||^{2}=1$. Thus, we have
\begin{equation}
||\phi_{n}||^{2}=i\int_{S}\left[\left(\partial_{\mu}\phi_{n}\right)\phi_{n}^{*}-\phi_{n}\left(\partial_{\mu}\phi_{n}^{*}\right)\right]\sqrt{-\hat{g}_{S}}n^{\mu}dS=1, \label{16}
\end{equation}
in which $S$ symbolizes the space-like hyper-surface throughout which the integration is performed and $n_{\mu}$ is a time-like vector. One can show that $\hat{g}_{S}=\frac{\hat{g}}{\hat{g}_{tt}}=-\frac{\sin^2{\theta}\Sigma^2}{r^4}\widetilde{C}^{2}_{\Omega}$ and that the vector $n^{\mu}$ is given by
\begin{equation}
\left(n^{\mu}\right)=r\sqrt{\frac{\widetilde{\Delta}}{\widetilde{A}}}\left[\frac{\widetilde{A}}{r^{2}\widetilde{\Delta}},0,0,-\frac{\widetilde{A}}{r\widetilde{\Delta}}\left(\Omega-\widetilde{\omega}_{d}\right)\right]. \label{17}
\end{equation}
With these definitions the integral (\ref{16}) yields
\begin{equation}
i\int_{S}\lbrace\left[\left(\partial_{t}\phi_{n}\right)\phi_{n}^{*}-\phi_{n}\left(\partial_{t}\phi_{n}^{*}\right)\right]n^{t}+\left[\left(\partial_{z}\phi_{n}\right)\phi_{n}^{*}-\phi_{n}\left(\partial_{z}\phi_{n}^{*}\right)\right]n^{z}\rbrace\sqrt{-\hat{g}_{S}}dS=1. \label{18}
\end{equation}
Taking into account the solutions (\ref{15}), after some algebra one arrives to the following expression for $N_n$
\begin{equation}
N_{n}^{2}=\frac{r^3}{\left(2\pi\right)^{2}L\omega_{n}\Sigma^2\sin{\theta}}\sqrt{\frac{\widetilde{A}}{\widetilde{\Delta}}}\widetilde{C}^{-3}_{\Omega}. \label{20}
\end{equation}

With the results obtained concerning the scalar normal modes, we can proceed to calculate the Casimir energy. This will be done in the
next section.

\section{Casimir energy}
With the completely defined normal modes we can calculate the Casimir energy. The operator energy density for the massless scalar field is
\begin{equation}
T_{tt}=\partial_{t}\phi\partial_{t}\phi^{*}-\frac{1}{2}\hat{g}_{tt}\left(\hat{g}^{\mu\nu}\partial_{\mu}\phi\partial_{\nu}\phi^{*}\right). \label{21}
\end{equation}
With the aid of Eqs. (\ref{06.1}) and (\ref{15}) Eq. (\ref{21}) becomes
\begin{equation}
T_{tt}\!=\!\frac{N_n^2}{2} \left\lbrace\!\left[\omega_n^2\!+\!\widetilde{C}^{-2}_{\Omega}\frac{r^2}{\Sigma}\!\left(k_x^2\!+\!\frac{\widetilde{\Delta}}{r^{2}}k_y^2\right)\!\right]\!\sin^2\!\left(\frac{n\pi}{L}z\right)\!+\!\frac{r^{2}}{\widetilde{\Delta} \sin^2{\theta}}\!\left(\frac{n\pi}{L}\right)^2\!\widetilde{C}^{-4}_{\Omega}\!\cos^2\!\left(\frac{n\pi}{L}z\right)\!\right\rbrace . \label{22}
\end{equation}
the expected value of this operator is given by
\begin{equation}
\left\langle0|T_{tt}|0\right\rangle=\sum_{n}\int \int T_{tt}dk_xdk_y. \label{23}
\end{equation}
The Casimir energy density is obtained through the expression
\begin{equation}
\langle\epsilon_{vac}\rangle=\frac{1}{V_{p}}\int_{S}\sqrt{-\hat{g}_{S}}v^{\mu}v^{\nu}\langle 0|T_{tt}|0 \rangle dS, \label{24}
 \end{equation}
where $V_P$ is the proper volume of the cavity, $V_p=\int\sqrt{-\hat{g}_{S}}dxdydz$. Thus
\begin{equation}
\langle\epsilon_{vac}\rangle=\frac{1}{V_{p}}
\int\int\int\sqrt{-\hat{g}_{S}}\hat{v}^{t}\hat{v}^{t}\sum_{n}\int \int T_{tt}dk_xdk_ydxdydz, \label{25}
 \end{equation}
in which $\hat{v}^{t}=\widetilde{C}_{\Omega}$ \cite{Sorge2}. The calculation of the integral (\ref{25}) with respect to $x$ and $y$ results in
\begin{equation}
\langle\epsilon_{vac}\rangle\!=\frac{\widetilde{C}^{2}_{\Omega}}{L}\sum_{n}\!\int\! \int \!\int\!\frac{N_n^2}{2}\! \left[\mathcal{F}_n\!\left(k_x,k_y\right)\!\sin^2\!\left(\frac{n\pi}{L}z\right)\!+\!\mathcal{G}_n\!\left(k_x,k_y\right)\!\cos^2\!\left(\frac{n\pi}{L}z\right)\!\right]\! dz dk_xdk_y, \label{26}
 \end{equation}
where
\begin{equation}
\mathcal{F}_n\left(k_x,k_y\right)=\left[\omega_n^2\!+\!\widetilde{C}^{-2}_{\Omega}\frac{r^2}{\Sigma}\!\left(k_x^2\!+\!\frac{\widetilde{\Delta}}{r^{2}}k_y^2\right)\!\right]\ \ \ \ \mbox{and}\ \ \ \ \mathcal{G}_n\left(k_x,k_y\right)=\frac{r^{2}}{\widetilde{\Delta} \sin^2{\theta}}\!\left(\frac{n\pi}{L}\right)^2\!\widetilde{C}^{-4}_{\Omega}. \label{27}
\end{equation}
The integral in $z$ in Eq. (\ref{26}) yields $\frac{1}{2}L$, so
\begin{equation}
\langle\epsilon_{vac}\rangle\!=\!\frac{\widetilde{C}^{2}_{\Omega}}{4}\sum_{n}\!\int\! \int N_n^2\left[\mathcal{F}_n\left(k_x,k_y\right)+ \mathcal{G}_n\left(k_x,k_y\right)\right]dk_xdk_y.\label{28}
 \end{equation}
Substituting the values of $\mathcal{F}_n$, $\mathcal{G}_n$ and $N_n ^2$ in the expression (\ref{28}) we have
 \begin{equation}
\langle\epsilon_{vac}\rangle\!=\frac{r^3}{2\left(2\pi\right)^{2}L\sin{\theta}\Sigma^2}\sqrt{\frac{\widetilde{A}}{\widetilde{\Delta}}}\widetilde{C}^{-1}_{\Omega}\!\sum_{n}\!\int\! \int \omega_n dk_xdk_y.\label{29}
 \end{equation}
Plugging Eq.(\ref{14}) into the above expression and making the transformations $\frac{\sqrt{\widetilde{\Delta}}}{r}k_{y}=\tilde{k}_{y}$ and $k_{x}^{2}+\tilde{k}_{y}^{2}=k_{\parallel}^{2}$, Eq. (\ref{29}) becomes
\begin{equation}
\langle\epsilon_{vac}\rangle\!=\frac{r^5}{L\sin{\theta}\Sigma^2}\sqrt{\frac{\widetilde{A}}{\widetilde{\Delta}^2\Sigma}}\widetilde{C}^{-2}_{\Omega}\!\sum_{n}\!\int\frac{1}{2}\frac{d^{2}k}{\left(2\pi\right)^{2}}\left[\left(\frac{n\pi}{s}\right)^{2}+k^{2}\right]^{\frac{1}{2}} , \label{30}
 \end{equation}
where $s^2=\frac{\widetilde{\Delta}\sin^2{\theta}\widetilde{C}^{2}_{\Omega}L^2}{\Sigma}$. The integral can be resolved by the Schwinger proper-time method and the sum can be calculated via Riemann zeta function regularization procedure (see \cite{Milton}), which result in
  \begin{equation}
\langle\epsilon_{vac}\rangle\!=-\frac{\pi^2}{1440L^4}\sqrt{\frac{\widetilde{A}}{\widetilde{\Delta}^5}}\frac{r^5}{\Sigma \widetilde{C}^{5}_{\Omega}\sin^4\theta}. \label{31}
 \end{equation}
Finally, knowing that the proper length of the cavity is $L_p=\widetilde{C}_{\Omega}\frac{\sin{\theta}\sqrt{\widetilde{\Delta}}}{r}L$, the Casimir energy of a cavity orbiting a massive spherical rotating body surrounded by quintessence is
  \begin{equation}
\langle\epsilon_{vac}\rangle\!=-\frac{\pi^2}{1440L_{p}^{4}}\sqrt{\frac{r^2}{\Sigma}}\left[1-\frac{\widetilde{A}^{2}}{\widetilde{\Delta} \Sigma ^2}\sin^2{\theta}\left(\Omega-\widetilde{\omega}_d\right)^2\right]^{\frac{1}{2}}. \label{32}
 \end{equation}
When $\theta=\frac{\pi}{2}$ (equatorial plane) and $\overline{\alpha}=0$, one obtains the result found in \cite{Sorge2}. An interesting case is the cavity at the north pole, whose Casimir energy is simply
  \begin{equation}
\langle\epsilon_{vac}\rangle\!=-\frac{\pi^2}{1440L_{p}^{4}}\sqrt{\frac{r^2}{r^2+a^2}} , \label{33}
 \end{equation}
where there is not any dependence on the quintessential matter, just on the angular momentum of the source. It is worth also noticing that, when one turns off the rotation of the source ($a=0$), one obtains the Casimir energy of the massless scalar field in Minkowsky vacuum.

Specializing for $\theta=\pi/2$ and $\Omega=0$ ({\it i.e.}, the cavity does not orbit the source), Eq. (\ref{32}) becomes
  \begin{equation}
\langle\epsilon_{vac}\rangle\!=-\frac{\pi^2}{1440L_{p}^{4}}\left[1-\frac{4M^2a^2}{\left(r^2+a^2-2Mr-\overline{\alpha} r^{1-3\omega_o}\right)r^2}\right]^{1/2} , \label{34}
 \end{equation}
From this expression, another remarkable result that can be obtained is considering $\omega_o=-1$, when we have the Kerr-de Sitter spacetime. In this case $\overline{\alpha}=\Lambda/3$, where $\Lambda$ is the cosmological constant. In Fig.2 bellow we depict the modulus of the ratio between the Casimir energy in the Kerr spacetime with quintessence and that one calculated in the flat space, $R$, given by
\begin{equation}\label{35}
R=\left[1-\frac{4M^2a^2}{\left(r^2+a^2-2Mr-\overline{\alpha} r^{1-3\omega_o}\right)r^2}\right]^{1/2},
\end{equation}
as a function of the radial coordinate $r$, in natural units. The values used for the other parameters are $M=10$, $a=4$, $\omega_o=-2/3$ and $\overline{\alpha}=0.001$ in those units.
\begin{figure}[h!]
\centering
\includegraphics[scale=0.9]{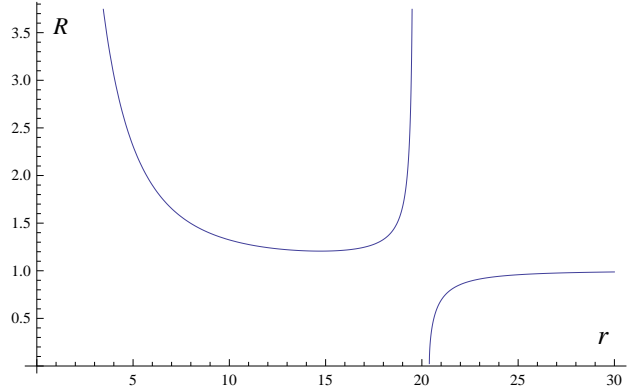}
\caption{Ratio $R$ between the Casimir energy in the Kerr spacetime with quintessence and that one in the flat spacetime, as a function of the radial coordinate $r$, in natural units.}
\end{figure}
The first one of the depicted disjunct curves represent the behavior of the Casimir energy between the first event horizon, at $r=0.83$ and the second one, at $r=19.2$. Notice the minimum at $r\thickapprox 15$. The second curve shows its behavior beyond the second horizon. Moreover, we can verify that the Kerr spacetime surrounded by quintessence presents a multi-horizon structure, in such a way that in the analyzed case there is a third horizon at about $r=980$, not shown in the graph. The ratio $R$ is equal to one in this region, presenting a jump (discontinuity) at the third horizon, beyond which $R=1$ again.

\section{Conclusions and Remarks}

We have calculated the Casimir energy of a massless scalar field in a cavity formed by nearby parallel plates orbiting a rotating spherical body surrounded by quintessence, according to the spacetime studied in \cite{Ghosh}. In this way we investigated the influence of the gravitational field on this energy, including the effects due to the spacetime dragging as well as to the quintessence, at zero temperature. The energy we have found depends on all the involved parameters, namely, the black hole mass, angular momentum and quintessence state parameter, for any radial coordinate and polar angle, with exception of the north pole, where the Casimir energy is not influenced by the quintessential matter. It is worth calling attention to the fact that at the equatorial plane, if the quintessence is canceled, the result obtained in \cite{Sorge2} is recovered, as expected.

The above results in considering the equatorial plane were depicted in a graph representing the behavior of the Casimir energy in the background spacetime under consideration and compared with the flat spacetime case, as a function of the radial coordinate, showing that it is compatible with the multi-horizon structure of this kind of black hole. Another result that we can point out here is obtaining limits on the quintessence parameters from the uncertainty in the current measurements of the Casimir effect, $(\langle\epsilon_{vac}\rangle-\langle\epsilon_{vac}^0\rangle)/\langle\epsilon_{vac}^0\rangle\sim 0.01$ \cite{Moste}, where $\langle\epsilon_{vac}^0\rangle$
is the Casimir effect calculated in the Minkowsky spacetime. Inserting the Earth data in Planck units in the expression (\ref{34}), we obtain the constraint $\overline{\alpha}\lesssim 5\times 10^{87}(4\times10^{41})^{3\omega_o-1}$. Thus, the constraints in $\overline{\alpha}$ ranges from $\overline{\alpha}\lesssim 6\times 10^{92} s^{-2}$, now in S.I. units, for $\omega_o=-1/3$, to $\overline{\alpha}\lesssim 4\times 10^{-8} s^{-2}$, for $\omega_o=-1$, which is compatible with the fact that, in this latter case, $\overline{\alpha}=\Lambda/3\thickapprox 10^{-35} s^{-2}$ according to measurements of the cosmological constant shown in \cite{Barrow}. This means that the concentration of dark energy around Earth can be much larger than the its average throughout the Universe, and in massive compact objects must be even more large.

As a future perspective we intend to study the thermal effects on the Casimir energy analyzed here.

\section{Acknowledgements}
We acknowledge the financial support by Conselho Nacional de Desenvolvimento Cientifico e Tecnl\'ogico(CNPq).

\end{document}